\documentclass[twocolumn,showpacs,preprintnumbers,amsmath,amssymb,prb]{revtex4}

\usepackage{graphicx}
\usepackage{dcolumn}
\usepackage{bm}

\begin{document}


\title{Theoretical study of the electronic states of hollandite vanadate K$_2$V$_8$O$_{16}$}

\author{S. Horiuchi}
\affiliation{Department of Physics, Chiba University, Inage-ku, Chiba 263-8522, Japan}
\author{T. Shirakawa}
\affiliation{Department of Physics, Chiba University, Inage-ku, Chiba 263-8522, Japan}
\author{Y. Ohta}
\affiliation{Department of Physics, Chiba University, Inage-ku, Chiba 263-8522, Japan}

\date{20 December 2007}

\begin{abstract}
We consider electronic properties of hollandite vanadate 
K$_2$V$_8$O$_{16}$, a one-dimensional zigzag-chain system 
of $t_{2g}$ orbitals in a mixed valent state.  
We first calculate the Madelung energy and obtain the relative 
stability of several charge-ordering patterns to determine 
the most stable one that is consistent with the observed 
superlattice structure.  
We then develop the strong-coupling perturbation theory to 
derive the effective spin-orbit Hamiltonian, starting 
from the triply-degenerate $t_{2g}$ orbitals in the VO$_6$ 
octahedral structure. 
We apply an exact-diagonalization technique on small clusters 
of this Hamiltonian and obtain the orbital-ordering pattern 
and spin structures in the ground state.  
We thereby discuss the electronic and magnetic properties 
of K$_2$V$_8$O$_{16}$ including predictions on the outcome 
of future experimental studies.  
\end{abstract}

\maketitle

\section{\label{sec:Intro}Introduction}

In the study of strongly correlated electron systems, 
vanadium oxide has been one of the central materials.  
In particular, the discovery of the phase transition 
associated with the reduction of the magnetic 
susceptibility in a mixed valence compound 
$\alpha'$-NaV$_2$O$_5$,\cite{isobe1,ohama} 
together with subsequent experimental and theoretical 
studies, has established the novel concept of 
the charge-ordering (CO) phase transition accompanied 
by the spin-singlet formation.  

Recently, Isobe {\it et al.}\cite{isobe} reported that, 
in hollandite vanadate K$_2$V$_8$O$_{16}$, a metal-insulator 
transition (MIT) occurs at $\sim$ 170~K, which is 
accompanied by the rapid reduction of the magnetic 
susceptibility.  
Below the transition temperature, a characteristic 
superlattice of $\sqrt{2}a\times \sqrt{2}a\times 2c$ is 
observed,\cite{isobe} whereby a possible CO phase 
transition accompanied by the spin-singlet formation 
was proposed.\cite{isobe} 

\begin{figure}[htbp]
\begin{center}
\resizebox{6.5cm}{!}{\includegraphics{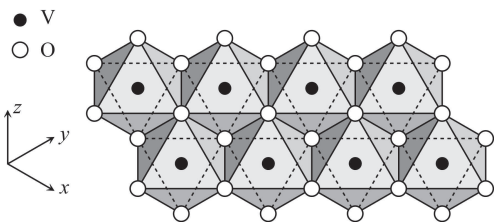}}\\
\caption{Schematic representation of the double string 
of edge-shared VO$_6$ octahedra in K$_2$V$_8$O$_{16}$.}	
\label{structure}
\end{center}		
\end{figure}

The crystal structure of this compound belongs to a group of 
hollandite-type phases and has a V$_{8}$O$_{16}$ framework 
composed of double strings of edge-shared VO$_6$ octahedra as 
shown in Fig.~\ref{structure}.
In the view of orbital physics, this system may be 
regarded as a one-dimensional (1D) version of LiVO$_2$ 
known as a possible orbital-ordering (OO) system of 
the $t_{2g}$ orbitals on the 2D triangular lattice 
of $S=1$ spins,\cite{pen,pen2} as in the case of 
a similar hollandite vanadate 
Bi$_x$V$_8$O$_{16}$.\cite{waki,shibata} 
The present system K$_2$V$_8$O$_{16}$ however has the 
average valence of V$^{3.75+}$ and thus is in the mixed 
valent state of V$^{3+}$:~V$^{4+}=3d^2:3d^1=1:3$, 
for which quite different electronic states are expected.  
Thus, the central issue in the present system is 
the mechanism of the MIT concerning how the highly 
frustrated spin, charge, and orbital degrees of freedom 
at high temperatures are relaxed by lowering temperatures 
and what type of orders is realized in the 
ground state.  

In this paper, we first examine the charge degrees of 
freedom of this system by calculating the Madelung energy; 
we obtain several CO patterns and their relative 
stability to determine the most stable one that is 
consistent with the characteristic superlattice 
structure observed in the low-temperature phase 
of the material. 
We then develop the strong-coupling perturbation theory 
starting from the triply-degenerate $t_{2g}$ orbitals in 
the VO$_6$ octahedral structure and derive the effective 
spin-orbit Hamiltonian in the approximation of neglecting 
the orbital fluctuations.  
We apply an exact-diagonalization technique on small 
clusters to this Hamiltonian and obtain the OO pattern 
and spin-spin correlation functions in the ground state.  
We thereby suggest that the state of local singlets of 
two $s=1/2$ spins coexisting with local high-spin 
clusters of $S=3/2$ should be of possible relevance with 
the observed electronic and magnetic states of 
K$_2$V$_8$O$_{16}$.  

This paper is organized as follows:  
In Sec.~II, we calculate the CO pattern within the ionic 
model and derive the effective spin-orbit Hamiltonian 
based on the perturbation theory.  
In Sec.~III, we calculate the orbital and spin structures 
of the derived effective Hamiltonian and compare the 
results with experiment.  Summary is given in Sec.~IV.  

\section{\label{sec:Theory}Effective Hamiltonian}

Here, we derive the effective spin-orbit 
Hamiltonian based on the strong-coupling perturbation 
theory, where the unperturbed state adopted is the 
lowest-energy CO state obtained from the ionic model 
in the limit of vanishing hopping parameters.  

\subsection{\label{sec:Madelung}Madelung energy and CO patterns} 

\begin{figure}[htbp]
\begin{center}
\resizebox{6cm}{!}{\includegraphics{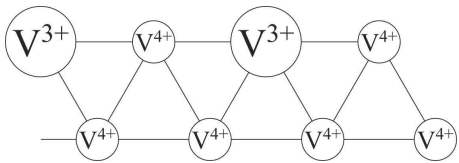}}\\
\caption{Most stable CO pattern in K$_2$V$_8$O$_{16}$ 
obtained from the Madelung energy calculations.  
Only a single zigzag chain of V ions in the material 
is illustrated.}
\label{CO}	
\end{center}		
\end{figure}

To determine the CO pattern in the ground state of 
K$_2$V$_8$O$_{16}$, we adopt the ionic model and 
calculate the Madelung energy,\cite{ziman} the 
electrostatic energy of an assembly of positive and 
negative point charges of the ions, which is known to be 
a good measure for the relative stability of the spatial 
distributions of $d$ electrons in transition-metal 
oxides\cite{kondo,ohta} as well as the CO patterns of some 
organic charge-transfer salts.\cite{mori}
We assume various $d$ electron distributions on the V ions 
and compare the Madelung energies where we use the 
room-temperature crystal structure reported for 
K$_2$V$_8$O$_{16}$.\cite{abriel} 
We first calculate the Madelung energy for the system 
with a hypothetical uniform electron distribution, 
where all the V ions are assumed to have the valence 
state of V$^{3.75+}$; we find the energy of 
$-3341.386$ eV/sc where sc is the unit cell of the observed 
$\sqrt{2}a\times\sqrt{2}a\times 2c$ superlattice.  
We then calculate the Madelung energies for various CO 
patterns with V$^{3+}$ or V$^{4+}$ and find that 
a number of CO patterns are more stable than the energy 
for the uniform electron distribution.  
The most stable CO pattern that is consistent with 
the superlattice structure observed in the low-temperature 
phase of this material is found to have the energy of 
$-3370.546$ eV/sc.  We point out that the obtained low-energy 
CO patterns commonly have the CO state illustrated in 
Fig.~\ref{CO}; i.e., the 1D zigzag chain consisting of 
V ions always has this CO state.  Therefore, we hereafter 
assume that this CO pattern is realized in K$_2$V$_8$O$_{16}$.  
We note that this CO pattern is different from the one 
proposed by Isobe {\it et al.} (see Fig.~4 of 
Ref.~\onlinecite{isobe}), which has a much higher Madelung 
energy $-3327.885$ eV/sc.  
Detailed discussions on the CO patterns are given in 
Ref.~\onlinecite{horiuchimaster}. 

\subsection{\label{sec:Effective}Perturbation theory}

\begin{figure}[htbp]
\begin{center}
\resizebox{6cm}{!}{\includegraphics{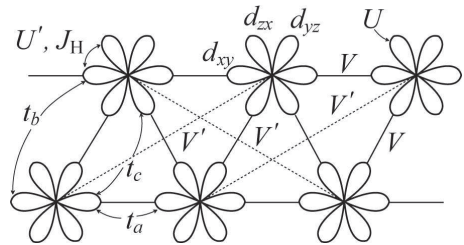}}\\
\caption{Schematic representation of the $t_{2g}$ 
orbitals on the 1D zigzag chain of K$_2$V$_8$O$_{16}$.  
Two of the four lobes for each of the three $t_{2g}$ 
orbitals are drawn.}	
\label{ham}
\end{center}		
\end{figure}

Here, we develop the strong-coupling perturbation theory 
starting from the ground state in the strong-coupling 
limit, i.e., the CO state shown in Fig.~\ref{CO}.  
Our starting high-energy Hamiltonian is of the 
following form:
\begin{eqnarray}
H&=&H_0+H_t \\
H_0&=&V\sum_{\langle ij\rangle}n_in_j+
V'\sum_{[ij]}n_in_j\nonumber \\
&&-J_{\rm H}\sum_{i\sigma\sigma',\alpha\ne \beta}
c_{i\alpha\sigma}^{\dagger}c_{i\beta\sigma'}^{\dagger}
c_{i\beta\sigma}c_{i\alpha\sigma'} \nonumber \\
&&+U\sum_{i\alpha}n_{i\alpha\uparrow}n_{i\alpha\downarrow}
+U'\sum_{i,\alpha\ne\beta}n_{i\alpha}n_{i\beta}\\
H_t&=&-\sum_{\langle i\alpha,j\beta\rangle,\sigma}
t_{i\alpha,j\beta}(c_{i\alpha\sigma}^{\dagger}c_{j\beta\sigma}+
{\rm H.c.})
\end{eqnarray}
where $c_{i\alpha\sigma}^{\dagger}\,(c_{i\alpha\sigma})$ 
is the creation (annihilation) operator of an electron 
at site $i$, orbital $\alpha$, and spin 
$\sigma=\uparrow,\downarrow$.  We define the number operators 
$n_{i\alpha\sigma}=c_{i\alpha\sigma}^{\dagger}c_{i\alpha\sigma}$, 
$n_{i\alpha}=n_{i\alpha\uparrow}+n_{i\alpha\downarrow}$, and 
$n_i=\sum_\alpha n_{i\alpha}$.  
$V$ and $V'$ are the intersite Coulomb repulsions 
between nearest-neighbor and next-nearest-neighbor sites, 
respectively.  
$J_{\rm H}$ is the Hund's rule coupling, and $U$ and $U'$ 
are the intra- and inter-orbital on-site Coulomb repulsions, 
respectively.  
We assume the relation $U'=U-2J_{\rm H}$ throughout the paper, 
which is valid in the atomic limit.  
$t_{i\alpha,j\beta}$ is the hopping parameter between 
the orbital $\alpha$ on site $i$ and orbital $\beta$ 
on site $j$ where $\alpha, \beta\in\{d_{xy},d_{yz},d_{zx}\}$ 
in the coordinate system shown in Fig.~\ref{structure}. 
We retain only the direct V-V hoppings between the $t_{2g}$ 
orbitals because the indirect hoppings via the O ions are 
rather small.\cite{pen}  
We then have the independent nearest-neighbor hopping 
parameters $t_a$, $t_b$, and $t_c$ as shown in Fig.~\ref{ham}.  

We assume that the ground state of our Hamiltonian $H_0$ 
should have the CO pattern shown in Fig.~\ref{CO}, where 
there are no doubly occupied orbitals and no V$^{2+}$ or 
V$^{5+}$ ionic states since these states are highly 
unrealistic in this material.  We therefore impose the 
condition
\begin{eqnarray}
U'-2V-J_{\rm H}>0.
\label{range}
\end{eqnarray}
We should also note that the unperturbed ($H_t=0$) states 
are spin- and orbital-degenerate of the degeneracy 
$M=3^N\cdot 3^{N/4}\cdot 2^{3N/4}$ where $N$ 
is the number of sites. 
This degeneracy is lifted by the perturbation processes. 
We develop the second-order perturbation 
calculation with respect to $t_{i\alpha,j\beta}$ 
assuming that $t_{i\alpha,j\beta}$ is much 
smaller than $U$, $U'$, $V$, and $V'$.  We thereby 
derive the effective spin-orbit Hamiltonian: 
\begin{eqnarray}
H_{\rm eff}&=&H_0-\sum_{\mu\mu'}\vert\mu\rangle
\sum_n\frac{\langle\mu\vert H_t\vert n\rangle
\langle n\vert H_t\vert\mu'\rangle}{E_n-E_0}\langle\mu'\vert
\end{eqnarray}
where $\vert\mu\rangle$ and $\vert\mu'\rangle$ 
($\mu,\mu'=1, \cdots, M$) are the $M$ independent 
eigenvectors of the ground state of $H_0$ and 
$\vert n\rangle$ is the $n$-th excited state of $H_0$. 
$E_n$ is the corresponding eigenenergy of $H_0$ 
($n=0$ denotes the ground state). 

To consider the real material, we should take into account 
the effect of distortion of the VO$_6$ octahedra more 
carefully because the degeneracy of the $t_{2g}$ orbitals 
can be lifted.  To evaluate the effect of distortion, 
we calculate the local symmetry of the Madelung site potential 
$\phi$ in the point charge model for the real material.  
Calculated result shows that the orbital $d_{xy}$ 
(see Fig.~\ref{ham}) is much less stable for the electron to 
sit on than the other orbitals $d_{yz}$ and $d_{zx}$ are; 
i.e. $\phi_{xy}>\phi_{yz}=\phi_{zx}$, where we also point out 
that the two orbitals $d_{yz}$ and $d_{zx}$ are exactly 
degenerate due to symmetry of the lattice.  
We therefore assume that the electrons do not occupy the 
$d_{xy}$ orbital in the ground state as well as in the 
perturbation processes.  A recent NMR experiment\cite{okai} 
seems to support this assumption.  

We introduce an approximation here; 
because the hopping parameters $t_{i\alpha,j\beta}$ 
take the values $t_a\gg t_b\simeq t_c$, we assume $t_b=t_c=0$ 
for simplicity as in Refs.~\onlinecite{pen} and 
\onlinecite{shibata}. 
This approximation means that the terms like 
$H_{ij}^{\rm eff}\propto t_at_c$ and 
$H_{ij}^{\rm eff}\propto t_bt_c$ are all neglected, 
retaining only the terms like $H_{ij}^{\rm eff}\propto t_a^2$ 
in the second-order processes.  
Note that the orbital fluctuations are completely suppressed 
in this approximation because only two orbitals connected with 
the {\it diagonal} hopping $t_a$ come out and no 
{\it off-diagonal} hopping terms appear in the effective 
Hamiltonian.  
We then obtain the effective spin-orbit Hamiltonian consisting 
of orbital-diagonal spin-subblocks with vanishing orbital 
off-diagonal blocks.  

\begin{figure}[htbp]
\begin{center}
\resizebox{7cm}{!}{\includegraphics{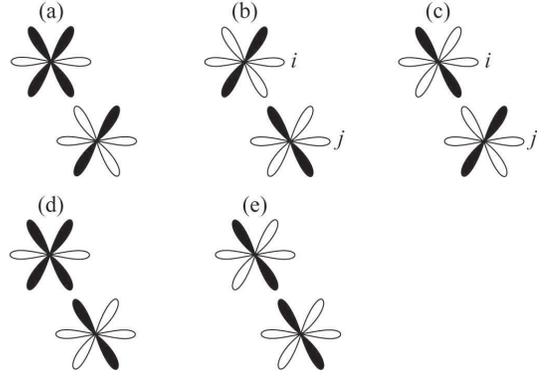}}\\
\caption{Schematic representation of five types of the 
bonds with different exchange interaction. 
Electrons are located in the shaded lobes. 
(a)~FM-1: the ferromagnetic bond with the process 
$d_i^2d_j^1\rightarrow d_i^1d_j^2\rightarrow d_i^2d_j^1$, 
(b)~FM-2: the ferromagnetic bond with the process 
$d_i^1d_j^1\rightarrow d_i^2d_j^0\rightarrow d_i^1d_j^1$, 
(c)~FM-3: the ferromagnetic bond with the process 
$d_i^1d_j^1\rightarrow d_i^0d_j^2\rightarrow d_i^1d_j^1$, 
(d)~AF-1: the antiferromagnetic bond with the process 
$d_i^2d_j^1\rightarrow d_i^3d_j^0$ or 
$d_i^1d_j^2\rightarrow d_i^2d_j^1$, and 
(e)~AF-2: the antiferromagnetic bond with the process 
$d_i^1d_j^1\rightarrow d_i^2d_j^0$ or $d_i^0d_j^2\rightarrow d_i^1d_j^1$. 
Here, we assume that the site $i$ denotes the site of 
the upper chain of Fig.~\ref{CO}, 
i.e., V$^{3+}$-V$^{4+}$-V$^{3+}$-V$^{4+}$ chain, 
and the site $j$ denotes the sites of the lower chain 
of Fig.~\ref{CO}, i.e., V$^{4+}$-V$^{4+}$-V$^{4+}$ chain.}
\label{eff}	
\end{center}		
\end{figure}

In the obtained effective spin-orbit Hamiltonian, we have 
five types of the bonds with different spin exchange 
interactions as shown in Fig.~\ref{eff}; 
three of them (denoted as FM-1, 2, 3) are the bonds with 
ferromagnetic exchange interaction due to double-exchange or 
Hund's rule coupling mechanisms 
and two of them (denoted as AF-1, 2) are the bonds with 
antiferromagnetic exchange interaction due to kinetic-exchange 
mechanism. 
Defining the spin-1 operator on site $i$ as $\bm{S}_i$ and 
spin-1/2 operator on site $i$ as $\bm{s}_i$, 
we have the following Hamiltonian for each bond shown in 
Fig.~\ref{eff}. \\
(a)~The bond FM-1 obtained with the process 
$d_i^2d_j^1\rightarrow d_i^1d_j^2\rightarrow d_i^2d_j^1$:
\begin{eqnarray}
H_{ij}^{\rm eff}&=&-2J\bm{S}_i\cdot\bm{s}_j+c\hat{1}\\
J&=&\frac{t_a^2}{4V'}-\frac{t_a^2}{4(V'+2J_{\rm H})}\nonumber \\
c&=&-\frac{3t_a^2}{4V'}-\frac{t_a^2}{4(V'+2J_{\rm H})}.\nonumber
\end{eqnarray}
(b)~The bond FM-2 obtained with the process 
$d_i^1d_j^1\rightarrow d_i^2d_j^0\rightarrow d_i^1d_j^1$:
\begin{eqnarray}
H_{ij}^{\rm eff}&=&-4J\bm{s}_i\cdot\bm{s}_j+c\hat{1}\\
J&=&\frac{t_a^2}{4(U'-V'-J_{\rm H})}-\frac{t_a^2}{4(U'-V'+J_{\rm H})}\nonumber \\
c&=&-\frac{3t_a^2}{4(U'-V'-J_{\rm H})}-\frac{t_a^2}{4(U'-V'+J_{\rm H})}.\nonumber
\end{eqnarray}
(c)~The bond FM-3 obtained with the process 
$d_i^1d_j^1\rightarrow d_i^0d_j^2\rightarrow d_i^1d_j^1$:
\begin{eqnarray}
H_{ij}^{\rm eff}&=&-4J\bm{s}_i\cdot\bm{s}_j+c\hat{1}\\
J&=&\frac{t_a^2}{4(U'-2V+V'-J_{\rm H})}-\frac{t_a^2}{4(U'-2V+V'+J_{\rm H})}\nonumber \\ 
c&=&-\frac{3t_a^2}{4(U'-2V+V'-J_{\rm H})}-\frac{t_a^2}{4(U'-2V+V'+J_{\rm H})}. \nonumber
\end{eqnarray}
(d)~The bond AF-1 obtained with the process 
$d_i^2d_j^1\rightarrow d_i^3d_j^0$ or $d_i^1d_j^2\rightarrow d_i^2d_j^1$:
\begin{eqnarray}
H_{ij}^{\rm eff}&=&2J\bm{S}_i\cdot\bm{s}_j+c\hat{1}\\
J&=&-c=\frac{t_a^2}{2(U+U'-2V-V')}+\frac{t_a^2}{2(U-U'+V'+J_{\rm H})}.\nonumber 
\end{eqnarray}
(e)~The bond AF-2 obtained with the process 
$d_i^1d_j^1\rightarrow d_i^2d_j^0$ or $d_i^0d_j^2\rightarrow d_i^1d_j^1$:
\begin{eqnarray}
H_{ij}^{\rm eff}&=&4J\bm{s}_i\cdot\bm{s}_j+c\hat{1}\\
J&=&-c=\frac{t_a^2}{2(U-V')}+\frac{t_a^2}{2(U-2V+V')}.\nonumber 
\end{eqnarray}
Here, $\hat{1}$ is the unit operator and we assume that 
the site $i$ denotes the site of the upper chain of Fig.~\ref{CO}, 
i.e., V$^{3+}$-V$^{4+}$-V$^{3+}$-V$^{4+}$ chain, and the site $j$ 
denotes the sites of the lower chain of Fig.~\ref{CO}, i.e., 
V$^{4+}$-V$^{4+}$-V$^{4+}$ chain.  

Thus, we obtain the effective spin-orbit Hamiltonian 
\begin{eqnarray}
H_{\rm eff}=\sum_{\langle ij\rangle}H_{ij}^{\rm eff} 
\label{eq:Heff}
\end{eqnarray}
where the sum runs over all the nearest-neighbor 
pairs of sites.  Note that this effective spin-orbit 
Hamiltonian has the form of block-diagonal in the 
spin$\otimes$orbit space; i.e., orbital off-diagonal 
blocks are all zero.  
In other words, we have several OO patterns for the 
$N$-site systems, and each of them, we have the spin 
Hamiltonian.  
By diagonalizing all the spin Hamiltonians, we can 
obtain the eigenstates of our effective spin-orbit 
Hamiltonian.  In particular, from the lowest-energy 
eigenstate obtained, we determine the OO pattern of 
the ground state in the parameter space.  

\section{\label{sec:Results}Results of calculation}

In this section, we calculate the orbital and spin structures 
in the ground state of the effective spin-orbit Hamiltonian 
Eq.~(\ref{eq:Heff}) by using the exact-diagonalization technique 
on small clusters, whereby we discuss its electronic 
and magnetic properties and compare them with experiment.  
We assume $U'=U-2J_{\rm H}$ throughout the calculations.  

\begin{figure}[tbhp]
\begin{center}
\resizebox{7cm}{!}{\includegraphics{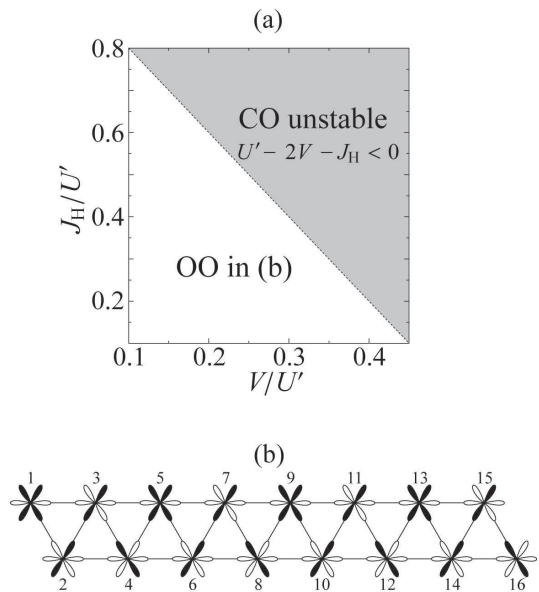}}\\
\caption{(a) Ground-state phase diagram of the 
effective spin-orbit Hamiltonian.  The CO pattern 
shown in Fig.~\ref{CO} is unstable in the shaded 
region. 
(b) Schematic representation of the calculated OO 
state in the lower-left region of the upper panel 
(a).}	
\label{orbital}
\end{center}		
\end{figure}

\subsection{\label{OO}Orbital ordering}

We here use the 16-site cluster with four $S=1$ spins and 
twelve $s=1/2$ spins (corresponding to the filling of 
20 electrons), where the spins are coupled with the 
exchange interactions derived in Sec.~II.  
The periodic boundary condition is used.  
Calculations are made for all possible OO patterns 
that are consistent with the two-fold periodicity along 
the $c$-axis observed in experiment; i.e., we assume 
the unit cell in the presence of the orbital ordering, 
which contains the 4 sites along the chain direction 
(e.g., sites $1-4$ in Fig.~\ref{orbital}~(b)).  
Hereafter we assume the relation $V'=0.6V$ estimated 
from the experimental interatomic distances between 
V ions.  

\begin{figure}[hbt]
\begin{center}
\resizebox{8cm}{!}{\includegraphics{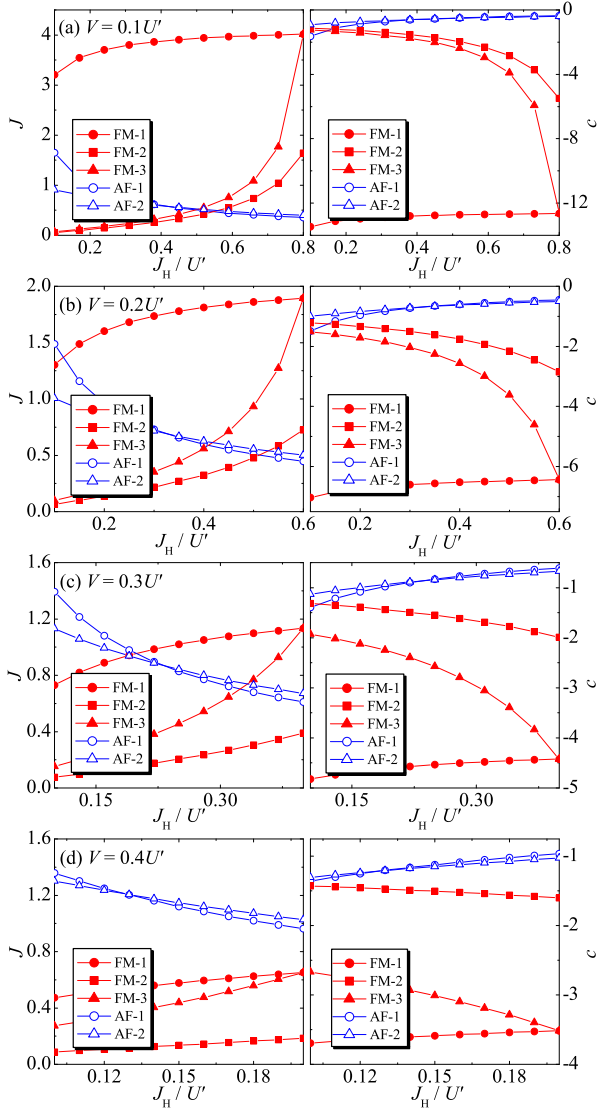}}\\
\caption{(Color online) Calculated values of the exchange 
coupling constant $J$ (left panels) and the constant 
$c$ (right panels) given in units of $t_a^2/U'$.}	
\label{Jc}
\end{center}		
\end{figure}
The calculated result is shown in Fig.~\ref{orbital}, 
where we find that the unique orbital state illustrated in 
Fig.~\ref{orbital}~(b) with the total-spin quantum number 
$S_{\rm tot}=0$ is realized in the entire 
parameter space shown in Fig.~\ref{orbital}~(a) unless 
the CO state is unstable.  The parameter space shown in 
Fig.~\ref{orbital}~(a) may be presumed to contain physically 
realistic values for vanadium oxide materials.\cite{miyasaka}  
Note that this orbital state is stabilized by maximizing 
the number of the FM-1 bonds as shown below. 

In Fig.~\ref{Jc}, we show calculated values of the 
exchange coupling constant $J$ and constant $c$ in the 
effective spin-orbit Hamiltonian Eqs.~(6)-(10).  
We find that the value of $c$ is predominantly 
lower for the FM-1 bond in the entire parameter 
space shown in Fig.~\ref{orbital}~(a).  
Thus, we understand that the OO pattern shown in 
Fig.~\ref{orbital}~(b) is stabilized by maximizing the 
number of this FM-1 bond.  
We should however note that, although in the small 
$V/U'$ region the value of $J$ for the FM-1 bond 
is larger than those of the other bonds, the values 
of $J$ for the antiferromagnetic bonds are much 
larger than those for the ferromagnetic bonds in the 
large $V/U'$ region.  In particular, a large value 
of $J$ for the AF-2 bond in the large $V/U'$ region 
is essential for the formation of the local spin-singlet 
of the two $s=1/2$ spins as shown below.  
We thus find that the spin structure can be quite 
different in the different regions in Fig.~\ref{orbital}~(a) 
although the unique OO pattern shown in 
Fig.~\ref{orbital}~(b) is stabilized.  

\subsection{\label{Spin}Spin correlations}

\begin{figure}[bth]
\begin{center}
\resizebox{8.6cm}{!}{\includegraphics{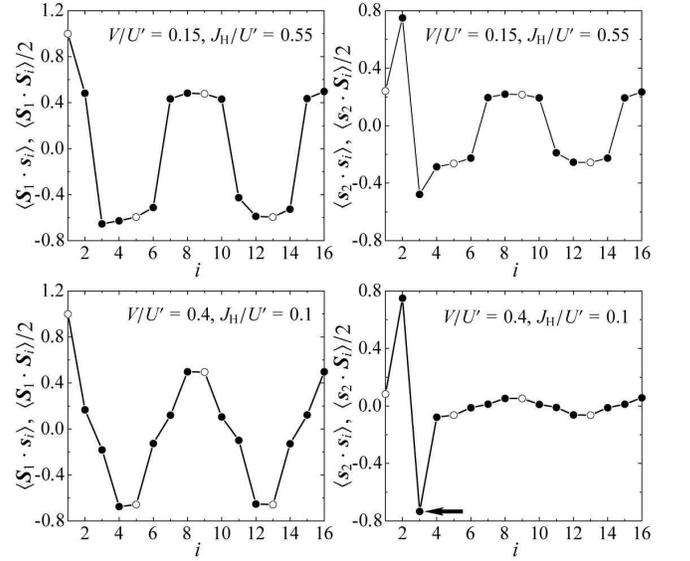}}\\
\caption{Spin-spin correlation functions calculated 
for the ground state of the 16-site cluster.  
The site index $i$ is defined in Fig.~\ref{orbital}~(b); 
the symbols $\circ$ and $\bullet$ indicate the sites 
with $S=1$ and $s=1/2$, respectively.  
Left and right panels represent the spin correlations 
from the site 1 ($S=1$) and site 2 ($s=1/2$), 
respectively. 
Upper and lower panels represent the spin correlations 
in the parameter regions $J_{\rm H}\gg V$ and 
$V\gg J_{\rm H}$, respectively.  
The arrow in the lower right panel indicates the 
formation of the local spin-singlet of the two $s=1/2$ 
spins.}	
\label{spincorrelation}
\end{center}
\end{figure}

We calculate the spin-spin correlation functions 
$\langle\bm{S}_i\cdot\bm{S}_j\rangle$, 
$\langle\bm{S}_i\cdot\bm{s}_j\rangle$, and 
$\langle\bm{s}_i\cdot\bm{s}_j\rangle$ for 
the ground state of the 16-site cluster of the 
effective spin-orbit Hamiltonian to discuss the spin 
structures of the system.  
The results are shown in Fig.~\ref{spincorrelation}, where we 
in particular show 
$\langle \bm{S}_1\cdot \bm{s}_i\rangle$ for $i\in s=1/2$ and 
$\langle \bm{S}_1\cdot\bm{S}_i\rangle/2$ for $i\in S=1$ 
in the left panels of Fig.~\ref{spincorrelation}, and 
$\langle\bm{s}_2\cdot\bm{s}_i\rangle$ for $i\in s=1/2$ and 
$\langle\bm{s}_2\cdot\bm{S}_i\rangle/2$ for $i\in S=1$ 
in the right panels of Fig.~\ref{spincorrelation}.  

We find in Fig.~\ref{spincorrelation} that the behavior 
of the spin correlations is very different between two 
regions of the OO phase in the white triangle area of 
Fig.~5~(a): i.e., the regions $J_{\rm H}\gg V$ and 
$V\gg J_{\rm H}$.  
This result suggests that, although the OO pattern is 
the same and the total spin $S_{\rm tot}=0$ in the two 
regions, two different spin structures can be realized 
depending on the parameter values.  

\begin{figure}[thb]
\begin{center}
\resizebox{8.2cm}{!}{\includegraphics{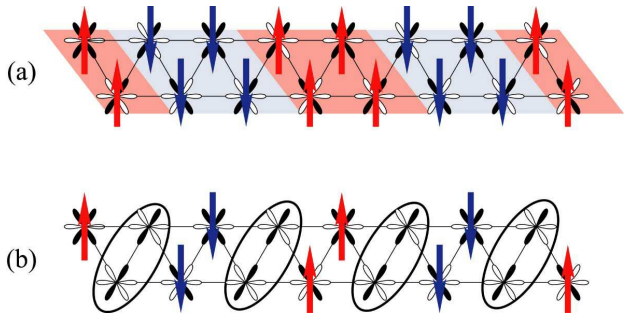}}\\
\caption{(Color online) Schematic representation of the 
spin structures.  
(a) The state of the antiferromagnetically fluctuating 
local high-spin clusters.  
(b) The state of the local spin-singlets of two $s=1/2$ 
spins (indicated by solid circles) coexisting with the 
local high-spin clusters of $S=3/2$.}
\label{spin}
\end{center}		
\end{figure}

In the region $J_{\rm H}\gg V$, we find the results that 
can roughly be described by the situation where the local 
high-spin clusters of $S=5/2$ formed by four V ions are 
coupled antiferromagnetically; we find the values 
$\langle\bm{S}_1\cdot\bm{s}_2\rangle=0.482$, 
$\langle\bm{S}_1\cdot\bm{s}_{15}\rangle=0.435$, and 
$\langle\bm{S}_1\cdot\bm{s}_{16}\rangle=0.498$, 
which should be $0.5$ if the formation of the 
high-spin cluster of $S=5/2$ were complete.  
We also find the values 
$\langle\bm{s}_2\cdot\bm{S}_1\rangle/2=0.241$, 
$\langle\bm{s}_2\cdot\bm{s}_{15}\rangle=0.194$, and 
$\langle\bm{s}_2\cdot\bm{s}_{16}\rangle=0.234$, 
which should be $0.25$ if the formation of the 
high-spin cluster of $S=5/2$ were complete.  
We also find the oscillations of $\pm 0.5$ ($\pm 0.25$) 
between two clusters of four spins in the upper-left 
(upper-right) panel of Fig.~7.  
Thus, our system in this parameter region can be regarded 
as the state of the antiferromagnetically fluctuating local 
high-spin clusters of $S=5/2$.  This state is illustrated 
schematically in Fig.~\ref{spin}~(a).  

In the region $V\gg J_{\rm H}$, we find the results that 
can roughly be described by the situation where the local 
spin-singlet states are formed between two $s=1/2$ spins; 
we find the value $\langle\bm{s}_2\cdot\bm{s}_3\rangle=-0.734$, 
which is only slightly larger than the value $-0.75$ that 
is expected when the spin-singlet formation is complete.  
We also note that the values of 
$\langle\bm{s}_2\cdot\bm{s}_i\rangle$ and 
$\langle\bm{s}_2\cdot\bm{S}_i\rangle$ are very small 
for all $i$ except $i=3$, which is consistent with the 
formation of the local spin-singlet state.  
We should however find the value 
$\langle\bm{S}_1\cdot\bm{s}_{16}\rangle=0.498$, 
which indicates the formation of the high-spin cluster 
of $S=3/2$ between the two spins at sites 1 and 16.  
We also note that the oscillation of $\pm 0.5$ 
between clusters of the two spins appears.  Thus, there remains 
antiferromagnetic correlations between the high-spin 
clusters of $S=3/2$.  This correlation occurs because the 
spin-singlet formation is strong but not perfect.  
Thus, our system in this parameter region can be regarded 
as the state of the local spin-singlets of two $s=1/2$ 
spins coexisting with the antiferromagnetically fluctuating 
local high-spin clusters of $S=3/2$.  This state is 
illustrated schematically in Fig.~\ref{spin}~(b).  

The spin structures thus obtained are compared with 
available experimental data in the next subsection.  

\subsection{\label{Expt}Comparison with experiment}

In the present theory, we have presumed that the MIT 
observed in K$_2$V$_8$O$_{16}$ is caused by the CO of 
$d$-electrons on the V ions.  We suppose this to be 
quite a natural interpretation because of the various 
features observed in experiment;\cite{isobe} e.g., 
the characteristic superlattice structure and lattice 
distortion observed below the transition temperature. 
We have also assumed that the electrons do not occupy 
the $d_{xy}$ orbital (see Fig.~\ref{ham}) as can be 
justified from the calculation of the Madelung site 
potential.  This is also consistent with the recent 
NMR measurement\cite{okai} of the symmetry axis of 
this material at room temperature.  These results 
first of all support the validity of our effective 
spin-orbit Hamiltonian derived in Sec.~II.  

Then, as for the magnetic aspects of K$_2$V$_8$O$_{16}$, 
the rapid reduction of the magnetic susceptibility observed 
below the transition temperature seems to suggest the 
opening of the spin gap.  It seems however that a small 
but finite value of the susceptibility remains finite 
at low temperatures as can be seen in Fig.~2 of 
Ref.~\onlinecite{isobe} even after subtracting the 
Curie term coming from the presence of impurities.  
This situation seems to suggest that the formation of 
nonmagnetic local spin-singlets occurs below the transition 
temperature but that there still remain interacting magnetic 
spins.\cite{isobe}  The broad nonmagnetic spectrum in 
the insulating region observed in a recent NMR 
experiment\cite{okai} might also be interesting in this 
respect.  
We therefore argue that these experimental situations 
are of possible relevance with the spin structure shown 
in Fig.~\ref{spin}~(b) where the local spin-singlets of 
two $s=1/2$ spins coexist with the antiferromagnetically 
interacting local high-spin clusters.  

\section{\label{sec:Summary}Summary}

We have studied the electronic and magnetic properties 
in hollandite vanadate K$_2$V$_8$O$_{16}$, a possible 
charge and orbital ordering system with the mixed valent 
state of V ions with $3d^2:3d^1=1:3$ and with the $t_{2g}$ 
orbitals of V ions aligned on the 1D zigzag chains.  
First, we have calculated the Madelung energy of the 
system and obtained the most stable CO pattern that 
is consistent with the superlattice structure observed 
in experiment.  
Then, by using the second-order perturbation theory 
starting from the triply-degenerate $t_{2g}$ orbitals 
in the VO$_6$ octahedral structure, we have derived 
the effective spin-orbit Hamiltonian.  
Here, we have evaluated the effect of distortions of 
the VO$_6$ octahedra from the local symmetry of the 
Madelung site potential, which justifies the assumption 
that the electrons do not occupy the $d_{xy}$ orbital.  
Within the approximation of neglecting the small 
off-diagonal hopping parameters, we have found that 
the Hamiltonian is block-diagonal with vanishing 
orbital off-diagonal sectors.  
We then have used the numerical exact-diagonalization 
technique on small clusters and have obtained the 
orbital-ordering pattern in the ground state.  
We have also calculate the spin-spin correlation functions 
and have found that, depending on the parameter values, 
either the state of the antiferromagnetically fluctuating 
local high-spin clusters or the state of the local 
spin-singlets of two $s=1/2$ spins coexisting with the 
local high-spin clusters of $S=3/2$ is realized.  
By comparing these results with available experimental 
data which are quite limited at present, we suggest that 
the latter state can be in agreement with the electronic 
ground state of hollandite vanadate K$_2$V$_8$O$_{16}$.  

Because our study presented here contains series of 
theoretical predictions on the outcome of future 
experimental studies, we hope that the present study 
will help one understand the nature of the charge, 
orbital, and spin degrees of freedom of this intriguing 
material.  

\begin{acknowledgments}
We would like to thank M. Isobe, M. Itoh, and K. Okai 
for useful discussions on the experimental aspects 
of K$_2$V$_8$O$_{16}$.  
This work was supported in part by 
Grants-in-Aid for Scientific Research 
(Nos. 18028008, 18043006, 18540338, and 19014004) 
from the Ministry of Education, Culture, Sports, 
Science and Technology of Japan.  
A part of computations was carried out at the 
Research Center for Computational Science, 
Okazaki Research Facilities, and the Institute 
for Solid State Physics, University of Tokyo.  
\end{acknowledgments}

\end{document}